\documentclass[final,5p,times,twocolumn]{elsarticle}

\usepackage{epsfig}
\usepackage{graphicx}
\usepackage{dcolumn}
\usepackage{amsmath}
\usepackage{amsfonts}
\usepackage{amssymb}

\journal{Physica C}

\begin{document}

\begin{frontmatter}



\title{Nearly isotropic upper critical fields in a SrFe$_{1.85}$Co$_{0.15}$As$_{2}$ single crystal}



\author [1]{Seunghyun Khim}
\author [2]{Jun Sung Kim}
\author [1]{Jae Wook Kim}
\author [3]{Suk Ho Lee}
\author [4]{F. F. Balakirev}
\author [5]{Yunkyu Bang}
\author [1]{Kee Hoon Kim}

\address [1]{FPRD, Department of Physics and Astronomy, Seoul National University, Seoul 151-742, Republic of Korea}
\address [2]{Department of Physics, Pohang University of Science and Technology, Pohang 790-784, Republic of Korea}
\address [3]{Optical Engineering Research Institute, Mokpo National University, Muan 534-729, Republic of Korea}
\address [4]{National High Magnetic Field Laboratory, Los Alamos National Laboratory, Los Alamos, NM 87545, USA}
\address [5]{Department of Physics, Chonnam National University, Kwangju 500-757, Republic of Korea}

\begin{abstract}
We study temperature dependent upper critical field $H_{\rm c2}$ of a SrFe$_{1.85}$Co$_{0.15}$As$_{2}$ single crystal  (\textit{T$_c$}=20.2 K) along \textit{ab}-plane and \textit{c}-axis through resistivity measurements up to 50 T. For the both crystalline directions, $H_{\rm c2}$ becomes nearly isotropic at zero temperature limit, reaching $\sim$ 48 T. The temperature dependence of the $H_{\rm c2}$ curves is explained by interplay between orbital and Pauli limiting behaviors combined with the two band effects.
\end{abstract}


\end{frontmatter}


\section{Introduction}
\label{}
The observation of high temperature superconductivity in iron-pnictides has triggered a surge of research activity in recent years to understand basic physical properties as well as pairing mechanism of the new superconducting system \cite{kamihara}. Upper critical field \textit{H$_{\rm c2}$} is one of key parameters to indicate underlying superconducting pairing strength and related electronic structure. Nearly isotropic \textit{H$_{\rm c2}$}(0) has been recently observed in both a hole-doped (Ba,K)Fe$_2$As$_2$ crystal and an electron doped SrFe$_{1.8}$Co$_{0.2}$As$_2$ thin film \cite{yuan, baily}. Moreover, as the parent material shows quasi-two dimensional Fermi surfaces, nearly isotropic \textit{H$_{\rm c2}$}(0) is unusual and in sharp contrast with the case of cuprates. These experimental findings raise a couple of puzzles regarding the physics behind: (1) is the phenomenon an intrinsic property of the whole iron-pnictides? (2) why are they insensitive to the change of electronic structure driven by hole- and electron-doping? (3) can the orbital limiting effect only, combined with multiband effects, explain the low temperature behavior?  To answer these questions, it is quite desirable to check the \textit{H$_{\rm c2}$} behaviors in various forms of iron-pnictides. In this study, we have determined \textit{H$_{\rm c2}$}(\textit{T}) curves up to 50 T in an electron doped SrFe$_{1.85}$Co$_{0.15}$As$_2$ single crystal. We found nearly isotropic \textit{H$_{\rm c2}$}(0) in the single crystal, consistent with the case of the Co 20 \%-doped thin film. Our results indicate that the isotropic \textit{H$_{\rm c2}$}(0) behavior is quite robust, independent of dopant types or amount as well as disorders in the `122' system.

\section{Experiments}
\label{}

Single crystals of SrFe$_{2-\textit{x}}$Co$_{\textit{x}}$As$_2$ were prepared by the Sn-flux method as described in detail in previous works \cite{jskim,hjkim}. X-ray diffraction indicated a single phase without noticeable impurities and energy-dispersive X-ray spectroscopy showed \textit{x} = 0.15. The magnetic field dependence of resistivity was measured at an isothermal condition by use of a short pulse magnet with a 20 ms decay time at NHMFL, Los Alamos National Laboratory. Different pieces showing same \textit{T$_c$} from one batch were measured in each field direction. Data were taken during the $H$ down-sweep using a lock-in amplifier at 100 kHz. The resistivity curves in rising and decreasing $H$ were well overlapped at overall temperatures except 3 - 7 K, indicating isothermal temperature conditions are achieved  during the pulse at overall temperatures except ones between 3 and 7 K.

\section{Result and discussion}
\label{}

Figure 1 shows the resistivity curves for $H$ parallel and perpendicular to \emph{c}-axis (denoted by \textit{H $\parallel$ c} and \textit{H $\perp$ c}, respectively). \textit{H$_{\rm c2}$} at a given temperature was determined by the field position at which 90 \% normal state resistivity is realized.  \textit{T$_{\rm c}$} was estimated to be 20.2 K with this criterion. Figure 2 depicts the evolution of \textit{H$_{\rm c2}$} for both \textit{H $\parallel$ c} and \textit{H $\perp$ c}. When \textit{H$_{\rm c2}$} curve was estimated with conditions of 50 \% or 10 \% normal state resistivity, the determined shapes were nearly identical to the one in Fig. 2 except nearly parallel shift of \textit{H$_{\rm c2}$} less than 5 T.
\begin{figure}
\begin{center}
\epsfig{file=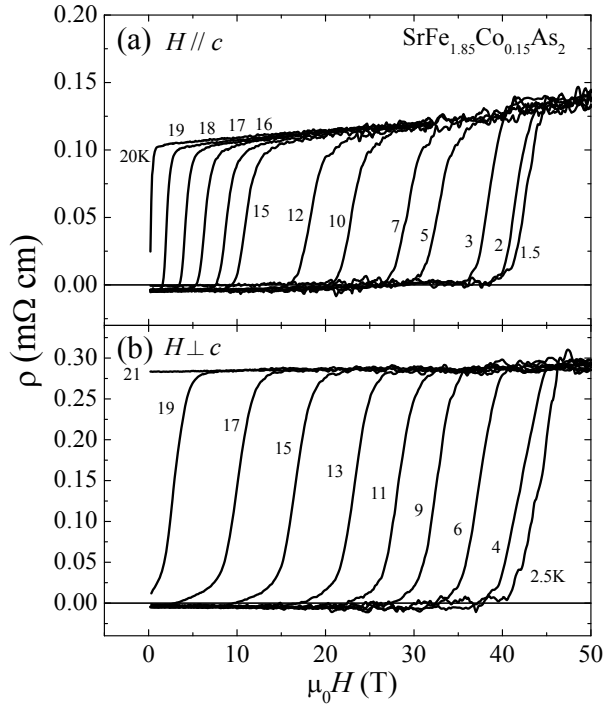,width=8 cm}
\end{center}
\caption{Resistivity measured under pulsed magnetic field up to 50 T at a fixed temperature. $H$ was applied (a) along \textit{c}-axis (\textit{H $\parallel$ c}) and (b) perpendicular to $c$-axis (\textit{H $\perp$ c}).}
\end{figure}

For \emph{H} $\parallel$ \emph{c}, \textit{H$_{\rm c2}$} is nearly linear in temperature, while for \emph{H} $\perp$ \emph{c}, \textit{H$_{\rm c2}$} has an upward curvature with temperature and shows saturation at low temperatures. Moreover, at overall temperatures, \textit{H$_{\rm c2}$} for \textit{H $\perp$ c} is higher than  for \textit{H $\parallel$ c}, showing more robust superconductivity for \textit{H $\perp$ c}. At the high temperature region near \textit{T$_c$}, the orbital limiting is thought to be a main mechanism for determining \textit{H$_{\rm c2}$}. The orbital limiting effect refers to opposite Lorentz forces on Cooper-paired electrons having reversed spatial momentum. From the initial slope near \textit{T$_c$}, we can estimate the orbital limiting field, \textit{H$_{c2}^{\rm orb}$}(0) in each direction. \textit{H$_{c2}$}(\textit T) slope near \textit{T$_{c }$} is 3.73 T/K and 2.45 T/K in \textit{H $\parallel$ c} and \textit{H $\perp$ c}, respectively. According to Werthamer-Helfand-Hohenberg (WHH) formula, i.e \textit{H$_{c2}^{\rm orb}$}(0) $\approx$ -0.69$\partial$\textit{H$_{c2}$}(\textit T)/$\partial$\textit{T}$\mid$$_{T=Tc}$\textit{T$_{c }$}, \textit{H$_{c2}^{\rm orb}$}(0) is 52.7 T and 34.6 T in each direction. The difference in \textit{H$_{c2}^{\rm orb}$}(0) should be related to the difference in quasiparticle effective mass (or superconducting coherence length) for each crystallographic direction. From the high temperature behavior, it is thus expected that the in-plance superconducting coherence length is larger than one along the \emph{c}-axis.

The dotted lines in Fig. 2 denote the temperature dependent WHH curve with $\alpha$ = $\lambda$ $_{so}$ = 0. According to the fit results, we note that the actual \textit{H$_{c2}$}(0) $\approx$ 48 T is larger than \textit{H$_{c2}$}(0) for \textit{H $\parallel$ c}, while it is smaller than \textit{H$_{c2}$}(0) for \textit{H $\perp$ c}. Unusually large \textit{H$_{c2}$}(0) beyond the predicted \textit{H$_{c2}^{\rm orb}$}(0) for \textit{H $\parallel$ c} is associated with its linear increase down to zero temperature. This linearly increasing \textit{H$_{c2}$} for \emph{H} $\parallel$ \emph{c} has been commonly observed in other `122' materials studied so far \cite{yuan,baily}. Indeed, the WHH formula cannot describe this linearly increasing behavior effectively as seen in the fitted curve. As studied recently \cite{baily}, the two band effects might have to be incorporated to properly explain the almost linearly increasing \textit{H$_{\rm c2}$} shape for \emph{H} $\parallel$ \emph{c}.

One conspicuous feature in Fig. 2 is the nearly isotropic \textit{H$_{\rm c2}$}(0) $\approx$ 48 T for both field directions. This is quite consistent with the results in an epitaxial thin-film \cite{baily}. When the ratio $\gamma$ between the two \textit{H$_{\rm c2}$} values is estimated from the interpolation of the curves in Fig. 2, $\gamma$ is around 1.7 near \textit{T$_{\rm c}$} and monotonically  decreases down to 1.05 at the zero temperature limit. Our observation implies that the nearly isotropic \textit{H$_{\rm c2}$} behavior is not sensitive to carrier doping and disorder. According to the fit results of the WHH formula for \emph{H} $\perp$ \emph{c}, the actual \textit{H$_{\rm c2}$}(0) $\approx$ 48 T is slightly, but clearly, less than the predicted \textit{H$_{c2}^{\rm orb}$}(0) = 52.7 T. This experimental fact suggests the need for the other mechanism, i.e., Pauli limiting to make the upper critical field saturate at low temperatures.

The Pauli limiting is based on the Zeeman effect on a spin-singlet Cooper pair so that it should be isotropic for any field direction. The Pauli limiting field \textit{H$_{p}$} of a BCS superconductor is estimated to be  1.84\textit{k$_{B}$T$_{c}$} = 37.5 T in our sample. This value is larger than the  orbital limiting field for \textit{H $\parallel$ c} but smaller than the \textit{H $\perp$ c} case. The corresponding Maki parameter $\alpha$ = $\sqrt{2}$\textit{H$_{c2}^{\rm orb}$}(0)/\textit{H$_{p}$} is $\alpha$ = 1.99 and 1.31 for \textit{H $\perp$ c} and \textit{H $\parallel$ c}, respectively.  This observation indicates that the \textit{H$_{\rm c2}$}(0) values in both directions are determined by the Pauli limiting effect while \textit{H$_{\rm c2}$}(0) along \textit{H $\perp$ c} is more severely limited and the \textit{H$_{\rm c2}$}(0) along \textit{H $\parallel$ c} is enhanced beyond the BCS prediction possibly due to spin-orbit coupling. Therefore, to fully explain the temperature dependence of \textit{H$_{\rm c2}$} for each field direction, one might have to consider the interplay between the Pauli and orbital limiting mechanism in the two-band system.
\begin{figure}
\begin{center}
\epsfig{file=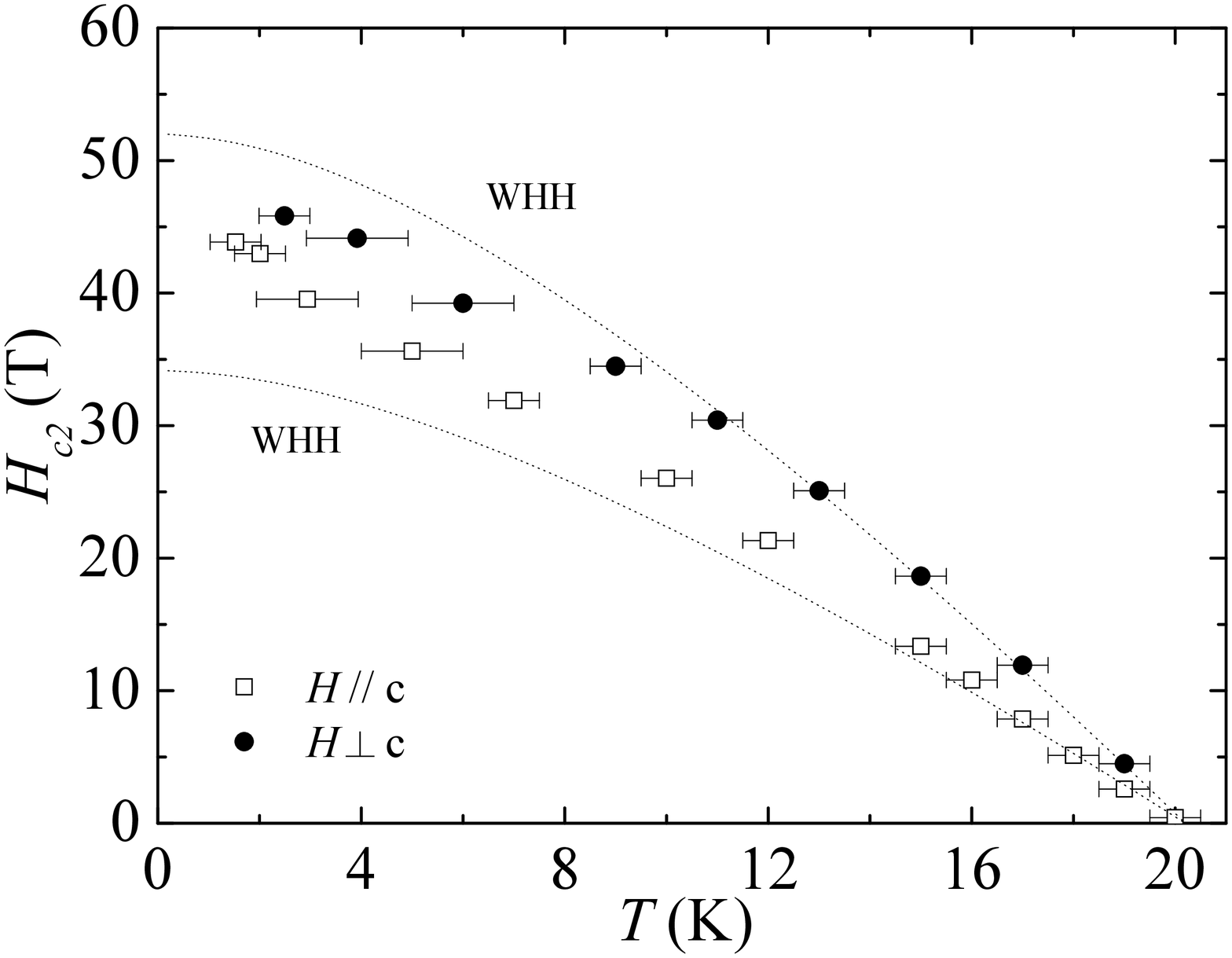,width=7 cm}
\end{center}
\caption{Temperature dependence of upper critical field \textit{H$_{\rm c2}$}. The dashed lines are the predictions of the WHH curve.}
\end{figure}

In summary, we find that a single crystal of SrFe$_{1.85}$Co$_{0.15}$As$_{2}$ shows nearly isotropic \textit{H$_{c2}$}(0) $\thickapprox$ 48 T in both crystallographic direction parallel and perpendicular to \emph{c}-axis, suggesting that the Pauli limiting effect should be also considered, in addition to the conventional orbital limiting effect in a multiband system.

This work is supported by Korean government through the NRF Korea (2009-0083512), NRL(M1060000238), and MOKE programs (Fundamental R$\&$D for Core Technology of Materials). KHK and SHK were supported by LG Yeonam Foundation and Seoul R$\&$BD (30932), respectively. Work at NHMFL was performed under the auspices of the NSF, the State of Florida, and the US DOE. We thank discussions with Prof. Hidenori Takagi (U. of Tokyo).



\end{document}